\documentclass[12pt]{article}
\usepackage[dvips]{graphicx}
\usepackage{epsfig}
\psfigdriver{dvips}
\begin{document}
\setcounter{page}{1}
\renewcommand{\thefootnote}{\fnsymbol{footnote}}
\renewcommand{\theenumi}{(\arabic{enumi})}
\renewcommand{\thefigure}{\arabic{figure}}
\renewcommand{\theequation}{\arabic{equation}}
\renewcommand{\arraystretch}{1.3}
\def\beq{\begin{equation}}
\def\eeq{\end{equation}}
\def\bea{\begin{eqnarray}}
\def\eea{\end{eqnarray}}
\def\bseq{\begin{subequations}}
\def\eseq{\end{subequations}}
\def\nn{\nonumber}
\def\dfrac{\displaystyle\frac}
\def\tfrac{\textstyle\frac}
\def\numt#1#2{#1 \times 10^{#2}}
\def\etal{{\it et al.}}
\def\etc{{\it etc.~}}
\def\ie{{\it i.e.,~}}
\def\eg{{\it e.g.~}}
\def\bs{\bigskip}
\def\ms{\medskip}
\def\ss{\smallskip}
\def\st{{\it s.t.,~}}
\def\id{{\mit I}}

\def\btiny{\begin{tiny}}
\def\etiny{\end{tiny}}
\def\bsc{\begin{scriptsize}}
\def\esc{\end{scriptsize}}
\def\bfoot{\begin{footnotesize}}
\def\efoot{\end{footnotesize}}
\def\bsm{\begin{small}}
\def\esm{\end{small}}
\def\bno{\begin{normalsize}}
\def\eno{\end{normalsize}}
\def\bla{\begin{large}}
\def\ela{\end{large}}
\def\bLa{\begin{Large}}
\def\eLa{\end{Large}}
\def\bLA{\begin{LARGE}}
\def\eLA{\end{LARGE}}
\def\bhu{\begin{huge}}
\def\ehu{\end{huge}}
\def\bHu{\begin{Huge}}
\def\eHu{\end{Huge}}

\def\bCe{\begin{center}}
\def\eCe{\end{center}}
\def\bFR{\begin{flushright}}
\def\eFR{\end{flushright}}
\def\bFL{\begin{flushleft}}
\def\eFL{\end{flushleft}}

\def\UL{\underline}

\def\PR#1#2#3{Phys. Rev. {\bf #1}, #2 (#3)}
\def\PRL#1#2#3{Phys. Pev. Lett. {\bf #1}, #2 (#3)}
\def\PL#1#2#3{Phys. Lett. {\bf #1}, #2 (#3)}
\def\NL#1#2#3{Nucl. Phys. {\bf #1}, #2 (#3)}
\def\NP#1#2#3{Nucl. Phys. {\bf #1}, #2 (#3)}
\def\PREP#1#2#3{Phys. Report {\bf #1}, #2 (#3)}
\def\Mod#1#2#3{Mod. Phys. Lett. {\bf #1}, #2 (#3)}
\def\PTP#1#2#3{Prog. Theor. Phys. {\bf #1}, #2 (#3)}
\def\EPJ#1#2#3{Eur. Phys. J. {\bf #1}, #2 (#3)}
\def\MPLA#1#2#3{Mod. Phys. Lett. {\bf A#1} (19#2) #3}
\def\PRD#1#2#3{Phys. Rev. {\bf D#1} (#2) #3}
\def\NPB#1#2#3{Nucl. Phys. {\bf B#1} (#2) #3}
\def\ZPC#1#2#3{Z. Phys. {\bf C#1} (#2) #3}
\def\EPJC#1#2#3{Eur. Phys. J. {\bf C#1} (#2) #3}
\def\PLB#1#2#3{Phys. Lett. {\bf B#1} (#2) #3}
\def\PRep#1#2#3{Phys. Rep. {\bf #1} (#2) #3}

\def\eqref#1{eq.(\ref{eqn:#1})}
\def\Eqref#1{Equation~(\ref{eqn:#1})}
\def\eqsref#1{eqs.(\ref{eqn:#1})}
\def\Eqsref#1{Equations~(\ref{eqn:#1})}
\def\eqvref#1{(\ref{eqn:#1})}
\def\Eqvref#1{(\ref{eqn:#1})}
\def\eqlab#1{\label{eqn:#1}}

\def\tbref#1{table~\ref{tbl:#1}}
\def\Tbref#1{Table~\ref{tbl:#1}}
\def\tbsref#1{tables~\ref{tbl:#1}}
\def\Tbsref#1{Tables~\ref{tbl:#1}}
\def\tbvref#1{\ref{tbl:#1}}
\def\Tbvref#1{\ref{tbl:#1}}
\def\tblab#1{\label{tbl:#1}}

\def\Fgref#1{Fig.\ref{fig:#1}}
\def\Figref#1{Figure~\ref{fig:#1}}
\def\Fgsref#1{Figs.~\ref{fig:#1}}
\def\Figsref#1{Figures~\ref{fig:#1}}
\def\fgvref#1{(\ref{fig:#1})}
\def\Fgvref#1{\ref{fig:#1}}
\def\Fglab#1{\label{fig:#1}}

\def\scref#1{section~\ref{sec:#1}}
\def\Scref#1{Section~\ref{sec:#1}}
\def\sclab#1{\label{sec:#1}}

\def\bmaT{\left(\begin{array}{ccc}}
\def\emaT{\end{array}\right)}
\def\bma{\left( \begin{array} }
\def\ema{\end{array} \right)}
\def\vev#1{\langle #1 \rangle}
\newcommand{\VEV}[1]{{\langle {#1} \rangle}}

\def\ov{\overline}
\def\wt{\widetilde}
\def\l{\left}
\def\r{\right}
\def\gsim{~{\rlap{\lower 3.5pt\hbox{$\mathchar\sim$}}\raise 1pt\hbox{$>$}}\,}
\def\lsim{~{\rlap{\lower 3.5pt\hbox{$\mathchar\sim$}}\raise 1pt\hbox{$<$}}\,}
\def\bmath#1{\mbox{\boldmath$#1$}}
\makeatletter
\newtoks\@stequation
\def\subequations{\refstepcounter{equation}%
  \edef\@savedequation{\the\c@equation}%
  \@stequation=\expandafter{\theequation}%   %only want \theequation
  \edef\@savedtheequation{\the\@stequation}% %expanded once
  \edef\oldtheequation{\theequation}%
  \setcounter{equation}{0}%
  \def\theequation{\oldtheequation\alph{equation}}}
\def\endsubequations{%
  \ifnum\c@equation < 2 \@warning{Only \the\c@equation\space subequation
    used in equation \@savedequation}\fi
  \setcounter{equation}{\@savedequation}%
  \@stequation=\expandafter{\@savedtheequation}%
  \edef\theequation{\the\@stequation}%
  \global\@ignoretrue}
\def\eqnarray{\stepcounter{equation}\let\@currentlabel\theequation
\global\@eqnswtrue\m@th
\global\@eqcnt\z@\tabskip\@centering\let\\\@eqncr
$$\halign to\displaywidth\bgroup\@eqnsel\hskip\@centering
     $\displaystyle\tabskip\z@{##}$&\global\@eqcnt\@ne
      \hfil$\;{##}\;$\hfil
     &\global\@eqcnt\tw@ $\displaystyle\tabskip\z@{##}$\hfil
   \tabskip\@centering&\llap{##}\tabskip\z@\cr}
\makeatother
\def\UMNS{{U_{_{\rm MNS}}}}
\def\cU#1#2{U_{#1}^{#2}}
\def\cUm#1#2{\wt{U}_{#1}^{#2}}
\def\satms#1{\sin^2\theta_{_{\rm ATM}}^{#1}}
\def\satmw#1{\sin^22\theta_{_{\rm ATM}}^{#1}}
\def\ssun#1{\sin^22\theta_{_{\rm SOL}}^{#1}}
\def\schz#1{\sin^22\theta_{_{\rm RCT}}^{#1}}
\def\matm#1{\delta m^{2~#1}_{_{\rm ATM}}}
\def\msun#1{\delta m^{2~#1}_{_{\rm SOL}}}
\def\dmns#1{\delta_{_{\rm MNS}}^{#1}}
\def\NBBn#1{{${\rm NBB}$($#1$GeV)}}
\def\NBBa#1{{$\overline{\rm NBB}$($#1$GeV)}}
\begin{titlepage}
\thispagestyle{empty}
\begin{flushright}
\begin{tabular}{l}
{KEK-TH-837}\\
{OCHA-PP-198}\\
{VPI-IPPAP-02-07}\\
{hep-ph/0208223}\\
\\
\end{tabular}
\end{flushright}
\baselineskip 24pt 
\begin{center}
{\Large\bf
Measuring the CP-violating phase
by a long base-line neutrino experiment
with Hyper-Kamiokande
}
\vspace{5mm}

\baselineskip 18pt 
\renewcommand{\thefootnote}{\fnsymbol{footnote}}
\setcounter{footnote}{0}

{Mayumi Aoki$^{1,2}$\footnote{{mayumi.aoki@kek.jp}},
 Kaoru Hagiwara$^1$,
 and
 Naotoshi Okamura$^{3}$\footnote{{nokamura@vt.edu}},
}\\
\bs
$^1${\it Theory Group, KEK, Tsukuba, Ibaraki 305-0801, Japan}\\
$^2${\it Department of Physics, Ochanomizu University, Tokyo 112-8610, Japan}\\
$^3${\it IPPAP, Physics Department, Virginia Tech. Blacksburg, VA 24061, USA}
\\
\end{center}
\begin{abstract}
We study the sensitivity of a long-base-line (LBL) experiment
with neutrino beams from the High Intensity Proton Accelerator (HIPA),
that delivers $10^{21}$ POT per year, and 
a proposed 1Mt water-$\check {\rm C}$erenkov detector,
Hyper-Kamiokande (HK) 295km away from the HIPA,
to the CP phase ($\dmns{}$) of the three-flavor lepton mixing matrix.
We examine a combination of the $\nu_\mu$ narrow-band beam (NBB)
at two different energies, $\vev{p_\pi}=2$, 3GeV, and
the $\overline\nu_\mu$ NBB at $\vev{p_\pi}=2$GeV.
By allocating one year each for the two $\nu_\mu^{}$ beams
and four years for the $\ov\nu_\mu^{}$ beam, we can efficiently measure
 the $\nu_\mu^{} \to \nu_e^{}$ and $\ov\nu_\mu^{} \to \ov\nu_e^{}$
transition probabilities, as well as the $\nu_\mu^{}$ and 
$\ov \nu_\mu^{}$ survival probabilities.
CP violation in the lepton sector can be established at 
4$\sigma$ (3$\sigma$) level if the MSW large-mixing-angle scenario
of the solar-neutrino deficit is realized,
$|\dmns{}|$ or
$|\dmns{}-180^{\circ}| >$ 30$^{\circ}$, and if
$4|U_{e3}^{}|^2 (1-|U_{e3}^{}|^2)\equiv \schz{} > 0.03$ (0.01).
The phase $\dmns{}$ is more difficult to constrain by this experiment
if there is little CP violation, $\dmns{}\sim 0^\circ$
or $180^\circ$, which can be distinguished
at 1$\sigma$ level
if $\schz{} \gsim 0.01$.
\end{abstract}
\bs
\bs
{
 \small 
 \begin{flushleft}
 \end{flushleft}
}
\begin{flushleft}
\end{flushleft}
\end{titlepage}
\newpage
\setcounter{page}{1}
\renewcommand{\thepage}{\arabic{page}}
\baselineskip 18pt 
\renewcommand{\thefootnote}{\fnsymbol{footnote}}
\setcounter{footnote}{0}
Neutrino oscillation experiment is one of the 
most attractive experiments
in the first quarter of 21st century.
Many experiments will measure precisely the model parameters
in the neutrino oscillations.
In this article,
we discuss the sensitivity of a long-base-line (LBL) 
experiment with
conventional neutrino beams to measure the CP phase in the
lepton sector.

The Super-Kamiokande (SK) collaboration showed that
the $\nu_\mu^{}$ created in the atmosphere
oscillates into $\nu_\tau^{}$ with almost maximal mixing
\cite{atm_tau}.
 The SNO collaboration reported that
the $\nu_e^{}$'s from the sun oscillate into the
other active neutrinos \cite{SNO}.
A consistent picture in the three active-neutrino framework 
is emerging.

In the three-neutrino-model, neutrino oscillations
depend on two mass-squared differences,
three mixing angles and one CP violating phase 
of the lepton-flavor mixing
(Maki-Nakagawa-Sakata (MNS) \cite{MNS}) matrix.
These parameters are constrained
by the solar and atmospheric neutrino observations.
 One of the mixing angles
and one of the mass-squared differences
are constrained by the atmospheric-neutrino observation,
which we may label \cite{H2B} as $\satms{}$ and
$\delta m^2_{_{\rm ATM}}$, respectively.
 The K2K experiment,
the ongoing LBL neutrino
oscillation experiment from KEK to SK,
constrains the same parameters \cite{K2K}.
 Their findings are consistent with the maximal mixing,
$\satmw{} \sim 1$ ($\satms{} \sim 0.5$) and
$\delta m^2_{_{\rm ATM}} \sim (2\sim4) \times 10^{-3}$(eV$^2$).
The solar-neutrino observations constrain 
another mixing angle
and the other mass-squared difference,
$\ssun{}$ and
$\delta m^2_{_{\rm SOL}}$, respectively.
Four possible solutions to the solar-neutrino deficit
problem \cite{solar} are found:
the MSW \cite{wolfenstein,MSW} large-mixing-angle (LMA) solution,
the MSW small-mixing-angle (SMA) solution,
the vacuum oscillation (VO) solution \cite{VO}, and
the MSW low-$\delta m^2$ (LOW) solution.
 The SK collaboration \cite{solar}
and the SNO collaboration \cite{SNO}
suggested 
that the MSW LMA solution is the most favorable solution among them,
for which $\ssun{}=0.7\sim0.9$ and $\msun{}=(3\sim 15) \times 10^{-5}$
eV$^2$.
 For the third mixing angle,
only the upper bound is obtained from
the reactor neutrino experiments.
CHOOZ \cite{CHOOZ} and Palo Verde \cite{PaloVarde} found 
$\sin^22\theta_{_{\rm RCT}} < 0.1$
for $\matm{} \sim \numt{3}{-3}$ eV$^2$.
No constraint on the CP phase
($\delta_{_{\rm MNS}}^{}$)
has been reported.

 Several future LBL neutrino-oscillation experiments
\cite{MINOS}-\cite{JHF2SK}
have been proposed to confirm the results of
these experiments and
to measure the neutrino oscillation parameters
more precisely.
 One of those experiments proposed in Japan 
makes use of the beam from
High Intensity Proton Accelerator (HIPA) \cite{HIPA_web}
and SK as the detector \cite{JHF2SK}.
 The facility HIPA \cite{HIPA_web} has a 50 GeV proton accelerator 
to be completed by the year 2007
in the site of JAERI (Japan Atomic Energy Research Institute),
as a joint project of KEK and JAERI.
 The proton beam of HIPA will deliver neutrino beams
of sub-GeV to several GeV range, whose intensity will be
two orders of magnitudes higher than that of the KEK PS beam for the K2K
experiment.
 The HIPA-to-SK experiment with $L$=295 km base-line length and
$\langle E_\nu \rangle \simeq 1$ GeV will measure
$\matm{}$ at about 3 $\%$ accuracy and
$\satms{}$ at about 1 $\%$ accuracy 
from the $\nu_\mu$ survival rate, while
$\nu_\mu^{}$-to-$\nu_e^{}$ oscillation can be discovered
if $\schz{} = 4|U_{e3}^{2}|(1-|U_{e3}^{2}|) \gsim$0.006
\cite{JHF2SK}.
 As a sequel to the HIPA-to-SK LBL experiment,
prospects of using the HIPA beam for a very long base-line
(VLBL) experiments with the base-line length of a few thousand
km have been studied \cite{H2B,BAND,walter}.
 Use of narrow-band high-energy neutrino beams
$(\vev{E_{\nu}^{}}=3 \sim 6$GeV$)$ 
and a 100kton-level water $\check {\rm C}$erenkov detector \cite{BAND}
will allow us to distinguish
the neutrino mass hierarchy (the sign of $m_3^2 - m_1^2$),
if $\sin^2 2\theta_{_{\rm RCT}}^{}\gsim0.03$ \cite{H2B}.
 If the LMA solution of the solar neutrino deficit is
chosen by the nature, we can further constrain the allowed region
of the $\delta_{_{\rm MNS}}^{}$ and
$\sin^22\theta_{_{\rm RCT}}^{}$ \cite{H2B}.
 However, because $\overline\nu_{\mu}^{}\to\overline\nu_e^{}$ appearance
is strongly suppressed by the matter effect at such high energies, 
the measurement is not sensitive to the CP violating
effects, $\sim \sin\delta_{_{\rm MNS}}^{}$.
 In this paper, we study the capability of an LBL experiment
between HIPA and Hyper-Kamiokande (HK), a megaton-level
water $\check {\rm C}$erenkov detector being proposed to be
built at the Kamioka site \cite{HK}.
 Here a combination of the shorter distance $(L=295$km$)$
and low $\nu$-energy $(\vev{E_\nu^{}}\sim 1$GeV$)$
makes the matter effect small, and the comparison of 
$\nu_\mu^{}\to\nu_e^{}$ and $\overline\nu_\mu^{}\to\overline\nu_e^{}$
appearance experiments is expected to have sensitivity
to the CP violation effects proportional to
$\sin \dmns{}$.

The MNS matrix of the three-neutrino model is
defined as
\begin{equation}
 \nu_\alpha^{}=
\sum_{i=1}^3 \l({{V}_{_{\rm MNS}}}\r)_{{\alpha}{i}}^{}
~{\nu_i^{}}
=
\sum_{i=1}^3 \l({{U}_{_{\rm MNS}}}\r)_{{\alpha}{i}}^{}
{\cal P}_{ii}
~{\nu_i^{}}
\,,
\eqlab{massdef}		
\end{equation}
where $\alpha=e,\mu,\tau$ are the lepton-flavor indices
and $\nu_i^{}~(i=1,2,3)$ denotes the neutrino
mass-eigenstates.
The $3\times3$ MNS matrix, $V_{\rm MNS}$,
has three mixing angles
and three phases in general for Majorana neutrinos.
In the above parameterization,
the two Majorana phases reside in the diagonal phase
matrix ${\cal P}$,
and the matrix $U$, which has three mixing angles and one phase,
can be parameterized in the same way as the CKM matrix \cite{CKM}.
 Because the present neutrino oscillation experiments
constrain directly the elements, $U_{e2}$, $U_{e3}$, and $U_{\mu 3}$,
we find it most convenient to adopt the parameterization \cite{HO1}
where these three matrix elements in the upper-right
corner of the $U$ matrix are chosen as the independent parameters.
 Without losing generality, we can take 
$U_{e2}$ and $U_{\mu 3}$ to be real and non-negative
while $U_{e3}$ is a complex number.
 All the other matrix elements of the $U$
are then determined by the unitary conditions \cite{HO1}. 

 The probability of finding the flavor-eigenstate $\beta$
at base-line length $L$ in the vacuum
from the original flavor-eigenstate $\alpha$
is given by
\bea
P_{\nu_\alpha \to \nu_\beta} &=&
\left| \cU{\beta 1}{} \cU{\alpha 1}{\ast}+\cU{\beta 2}{}
  e^{-i\Delta_{12}}
  \cU{\alpha 2}{\ast} 
+\cU{\beta 3}{}
  e^{-i\Delta_{13}}
  \cU{\alpha 3}{\ast}
\right|^2\,,
\eqlab{P_ex_0}
\eea
where
\begin{equation}
\Delta_{ij} \equiv \dfrac{m_j^2 - m_i^2}{2E_\nu}L
\simeq 2.534 \dfrac{\delta m_{ij}^2 ({\rm eV}^2)}{E_\nu({\rm GeV})}L({\rm km}
)
\end{equation}
satisfy
$\Delta_{12}^{}+\Delta_{23}^{}+\Delta_{31}^{}=
(\delta m^2_{12}+\delta m^2_{23}+\delta m^2_{31})(L/2E_\nu^{})=0$.
The two independent mass-squared differences
are identified with the two ``measured'' ones,
as follows;
\begin{equation}
 \delta m^2_{_{\rm SOL}} = \l|\delta {m}^2_{12}\r| \ll
\l|\delta {m}^2_{13}\r|  = \delta m^2_{_{\rm ATM}}\,.
\eqlab{relation_MSD}
\end{equation}
With the above identification,
the MNS matrix elements are constrained
by the observed survival probabilities,
$P_{\nu_\mu^{} \to \nu_\mu^{}}$
from the atmospheric neutrinos \cite{atm},
$P_{\overline{\nu}_e^{} \to \overline{\nu}_e^{}}$
from the reactor anti-neutrinos \cite{CHOOZ, PaloVarde},
and
$P_{{\nu}_e^{} \to {\nu}_e^{}}$
from the solar neutrinos \cite{solar}.
 The four independent parameters of the MNS matrix
are then related to the observed
oscillation amplitudes as
\bseq
\begin{eqnarray}
{{\l|U_{e 3}\r|^2}} &=&
\left(1 - \sqrt{ 1-\sin^2 2\theta_{_{\rm RCT}}}\right)
\scalebox{1.3}{$/^{^{}}$}2\,,
\eqlab{def_Ue3}\\
\l(U_{\mu 3}\r)^2&\equiv&
\satms{}
=\left(1 \pm \sqrt{ 1-\sin^2 2\theta_{_{\rm ATM}}}\right)
\scalebox{1.3}{$/^{^{}}$}2\,,
\eqlab{def_Um3}\\
\l(U_{e 2}^2\r)^2 &=&
\left(1 - { \l|U_{e3}\r|^2} -
\sqrt{\l(1 - { \l|U_{e3}\r|^2}\r)^2
-\sin^2 2\theta_{_{\rm SOL}}}\right)
\scalebox{1.7}{$/^{^{}}$}2\,, 
\eqlab{def_Ue2}\\
arg\l({ U_{e 3}}\r) &=& -{\delta_{_{\rm MNS}}^{}}\,.
\eqlab{def_delta}
\end{eqnarray}
\eqlab{def_U}
\eseq
\noindent
The CP phase of the MNS matrix, $\dmns{}$, is not constrained. 
The solution \eqref{def_Ue2} follows from our convention
\cite{H2B}, $U_{e1}>U_{e2}$, which defines the
mass-eigenstate $\nu_1^{}$.
In this convention,
there are four mass hierarchy cases corresponding to the
sign of $\delta m^2_{ij}$;
I $(\delta m^2_{13} > \delta m^2_{12} > 0)$,
II $(\delta m^2_{13} > 0 > \delta m^2_{12})$,
III $(\delta m^2_{12} > 0 > \delta m^2_{13})$, and
IV $(0>\delta m^2_{12} > \delta m^2_{13})$
\cite{H2B}.
 If the MSW effect is relevant for
the solar neutrino oscillation, then the neutrino mass hierarchy cases
II and IV are not favored.
 When $\satmw{}\neq 1$, there is an additional twofold
ambiguity in the determination of $U_{\mu 3}$ in \eqref{def_Um3}.
In order to avoid the ambiguity, we adopt the $U_{\mu 3}$
element itself, or equivalently $\satms{}$
defined in \eqref{def_Um3},
as an independent parameter of the MNS matrix.
Summing up, we parametrize the three-flavor neutrino oscillation
parameters in terms of the 5 observed (constrained) parameters
$\matm{}$, $\msun{}$, $\satms{}$, $\ssun{}$, $\schz{}$ and
one CP-violating phase $\dmns{}$, for four hierarchy cases.

Neutrino-flavor oscillation inside of the matter is governed by the
Schr{\" o}dinger equation
\begin{eqnarray}
i\frac{\partial}{\partial t}
\bma{c}
\nu_e \\
\nu_\mu \\
\nu_\tau\\
\ema
=
H
\bma{c}
\nu_e \\
\nu_\mu \\
\nu_\tau\\
\ema
=\frac{1}{2E_\nu}
\left[
{H_0} + 
\bmaT
a & 0 & 0 \\
0 & 0 & 0 \\
0 & 0 & 0
\emaT
\right]
\bma{c}
\nu_e \\
\nu_\mu \\
\nu_\tau\\
\ema\,,
\end{eqnarray}
where $H_0$ is the Hamiltonian in the vacuum
and $a$ is the matter effect term \cite{wolfenstein}
\begin{eqnarray}
a=2\sqrt{2} G_F^{} n_e^{} E_\nu^{} 
={7.56}\times 10^{-5}({\rm eV}^2)\left(\frac{\rho}{\rm g/cm^{3}}\right)
  \left(\frac{E_\nu}{\rm GeV}\right)\,.
\eqlab{matter_a}
\end{eqnarray}
Here
$n_e$ is the electron density of the matter,
$E_\nu^{}$ is the neutrino energy,
$G_F$ is the Fermi constant,
and $\rho$ is the matter density.
In our analysis,
we assume for brevity that the density of the earth's crust 
relevant for the LBL experiment, between HIPA
and HK is a constant,
$\rho=3$, with an overall uncertainty of $\Delta \rho=0.1$;
\begin{equation}
\rho ~({\rm g/cm}^3)=3.0\pm 0.1\,. 
\eqlab{def_err_mat}
\end{equation}
The Hamiltonian is diagonalized as
\begin{equation}
{H} = \dfrac{1}{2E_\nu}
{{\cUm{}{}}}
\bmaT
 {\lambda_1} & 0 & 0 \\
 0 & {\lambda_2} & 0 \\
 0 & 0 & {\lambda_3}
\emaT
{{\cUm{}{\dagger}}}
,
\eqlab{H_w_mat}
\end{equation}
by the MNS matrix in the matter $\wt U$.
The neutrino-flavor oscillation probabilities in the matter 
\begin{equation}
 P_{\nu_\alpha \to \nu_\beta} =
\l|\cUm{\beta 1}{} \cUm{\alpha 1}{\ast}
+\cUm{\beta 2}{}
  e^{-i{{\wt{\Delta}_{12}}}}
  \cUm{\alpha 2}{\ast}
+\cUm{\beta 3}{}
  e^{-i{{\wt{\Delta}_{13}}}}
  \cUm{\alpha 3}{\ast}
\r|^2\,,
\end{equation}
takes the same
form as those in the vacuum,
with
$\wt{\Delta}_{ij} = 
({{\lambda_j}-{\lambda_i}})L/{2E_\nu}
$,
if the matter density can be approximated by a constant throughout
the base-line.
Because the effective matter potential for anti-neutrinos has the
opposite sign with the same magnitude,
the total Hamiltonian $\overline H$ governing the anti-neutrino
oscillation in the matter is obtained from $H$ as follows \cite{H2B},
\begin{equation}
 \overline H\left(\delta m_{12}^2,\delta m_{13}^2\right)
=
 -H^{\ast}\left(-\delta m_{12}^2,-\delta m_{13}^2\right)\,.
\eqlab{def_antiH}
\end{equation}

 We make the following simple treatments
in estimating the signals and the backgrounds
in our analysis.
\begin{itemize}
\item 
We assume a 1 Mega-ton water $\check {\rm C}$erenkov detector,
which is capable of distinguishing between
$e^{\pm}$ CC events and $\mu^{\pm}$ CC events,
but cannot distinguish their charges.
\item
We do not require capability of the detector
to reconstruct the neutrino energy.
\end{itemize}
Although the water $\check {\rm C}$erenkov detector has the capability
of measuring the energy of the produced
$\mu^{\rm}$ and $e^{\rm}$ as well as
a part of hadronic activities,
we do not make use of these information
in this analysis.
We only use the total numbers of the produced $\mu^{\pm}$ and $e^{\pm}$
events from $\nu_\mu^{}$ or $\overline\nu_\mu^{}$ narrow-band-beams
(NBB).
 The NBBs from HIPA deliver $10^{21}$ protons on target (POT) in a typical
1 year operation, corresponding to about 100 days
of operation with the design intensity \cite{HIPA_web}.
Details of the NBB's used for this study are available from the
web-page \cite{NBB}.

In the following discussion,
we examine $\nu_{\mu}^{}$ NBB's with the mean
$\pi$ momentum $\vev{p_\pi}=2$GeV 
(\NBBn2) and $\vev{p_\pi}=3$GeV (\NBBn3),
and $\overline\nu_{\mu}^{}$ NBB with $\vev{p_\pi}=2$GeV,
(\NBBa2).
For our input (`true') value of $\matm{}=\numt{3.5}{-3}$ eV$^2$,
the probability $P_{\nu_\mu \to \nu_e}$ has a broad peak at
${E_\nu}\sim 1$GeV.
\NBBn2 and \NBBa2 are chosen to maximize the transition probability,
since $\vev{E_\nu}\simeq \vev{p_\pi}/2$.
Because $P_{\nu_\mu^{} \to \nu_e^{}}$
does not change much in the range
${E_\nu^{}}\simeq 0.6 \sim 1.2$ GeV,
our results do not depend strongly on
the true value of the $\matm{}$:
as long as it stays in the range
$(2 \sim 5) \times 10^{-3}$ eV$^2$ \cite{H2B}.

 The signals in this analysis are the numbers of
$\nu_\mu^{}$ and $\nu_e^{}$ CC events from \NBBn{2,3}
and those of the 
$\overline \nu_\mu^{}$ and $\overline \nu_e^{}$
CC events from \NBBa2.
These are calculated as
\bseq
\begin{eqnarray}
N_l^{}(\nu_\mu^{};\vev{p_\pi^{}})  
&=& MN_A\int_{0}^{10~\mbox{\tiny GeV}}
dE_\nu
\Phi_{\nu_\mu^{}}(E_\nu^{};\vev{p_\pi^{}}) 
P_{\nu_\mu \to \nu_\ell^{}}(E_\nu^{})
\sigma^{\rm CC}_{\nu_\ell^{}}(E_\nu^{})
\eqlab{event_n}\,,\\
\overline{N}_{\overline{l}}^{}(\overline{\nu}_\mu^{};\vev{p_\pi^{}})  
&=& MN_A\int_{0}^{10~\mbox{\tiny GeV}}
dE_{\overline{\nu}}
\overline{\Phi}_{\overline{\nu}_\mu^{}}(E_{\overline{\nu}}^{};\vev{p_\pi^{}}) 
P_{\overline{\nu}_\mu \to \overline{\nu}_{\ell}^{}}(E_{\overline{\nu}}^{})
\sigma^{\rm CC}_{\overline{\nu}_{\ell}^{}}(E_{\overline{\nu^{}}})
\eqlab{event_a}\,,
\end{eqnarray}
\eseq
\noindent
for $l = e$ or $\mu$, where 
$M$ is the mass of detector (1Mega-ton),
$N_A=\numt{6.017}{23}$ is the Avogadro number,
$\Phi_{\nu_\mu^{}}(E_\nu^{};\vev{p_\pi^{}})$ and
$\overline{\Phi}_{\overline{\nu}_\mu^{}}(E_{\overline{\nu^{}}};\vev{p_\pi^{}})$
are the flux
of $\nu_\mu^{}$ in \NBBn{\vev{p_\pi^{}}}
and $\overline{\nu}_\mu^{}$ in \NBBa{\vev{p_\pi^{}}},
respectively.
The flux is negligibly small at $E_{\nu}^{}>10$GeV for the NBB's
used in our analysis.
The cross sections are obtained by assuming a pure water target
\cite{cross}.
\begin{table}[t]
\begin{center}
\begin{small}
\begin{tabular}{|c|c||c|c||c|c||c|c|}
\hline
 &  
 & \multicolumn{2}{c||}{NBB $(\vev{p_\pi^{}}=2$GeV$)$}
 & \multicolumn{2}{c||}{NBB $(\vev{p_\pi^{}}=3$GeV$)$}
 & \multicolumn{2}{c|}{$\overline{\rm NBB}$$(\vev{p_\pi^{}}=2$GeV$)$}\\
\cline{3-8}
  $\schz{}$
& ${\dmns{}}^{}$ 
& $N_\mu^{^{}}$
& $N_e^{^{}}  $
& $N_\mu^{^{}}$
& $N_e^{^{}}  $
& $\overline{N^{}}_{\overline{\mu}}$%(\overline{\nu}_\mu;2{\rm GeV})$
& $\overline{N^{}}_{\overline{e}}  $%(\overline{\nu}_\mu;2{\rm GeV})$
\\
\hline
\hline
0.06&   $0^\circ$ 
& $\numt{5.0}{3}$ & $\numt{8.5}{2}$ 
& $\numt{1.6}{4}$ & $\numt{1.1}{3}$ 
& $\numt{1.6}{3}$ & $\numt{2.2}{2}$ \\
\hline
    &  $90^\circ$ 
& $\numt{5.1}{3}$ & $\numt{5.9}{2}$ 
& $\numt{1.6}{4}$ & $\numt{8.0}{2}$
& $\numt{1.6}{3}$ & $\numt{2.8}{2}$ \\
\hline
    & $180^\circ$ 
& $\numt{5.1}{3}$ & $\numt{7.9}{2}$ 
& $\numt{1.6}{4}$ & $\numt{9.1}{2}$
& $\numt{1.6}{3}$ & $\numt{2.0}{2}$ \\
\hline
    & $270^\circ$ 
& $\numt{5.1}{3}$ & $\numt{1.1}{3}$ 
& $\numt{1.6}{4}$ & $\numt{1.2}{3}$
& $\numt{1.6}{3}$ & $\numt{1.5}{2}$ \\
\hline
\hline
0.01&   $0^\circ$ 
& $\numt{5.1}{3}$ & $\numt{1.7}{2}$ 
& $\numt{1.6}{4}$ & $\numt{2.3}{2}$
& $\numt{1.6}{3}$ & $\numt{4.5}{1}$ \\
\hline
    &  $90^\circ$ 
& $\numt{5.1}{3}$ & $\numt{6.2}{1}$ 
& $\numt{1.6}{4}$ & $\numt{9.5}{1}$
& $\numt{1.6}{3}$ & $\numt{6.7}{1}$ \\
\hline
    & $180^\circ$  
& $\numt{5.1}{3}$ & $\numt{1.4}{2}$ 
& $\numt{1.6}{4}$ & $\numt{1.4}{2}$
& $\numt{1.6}{3}$ & $\numt{3.7}{1}$ \\
\hline
    & $270^\circ$  
& $\numt{5.1}{3}$ & $\numt{2.5}{2}$
& $\numt{1.6}{4}$ & $\numt{2.7}{2}$
& $\numt{1.6}{3}$ & $\numt{1.6}{1}$ \\
\hline
\end{tabular}
\end{small}
\end{center}
\caption{{%
Expected number of CC signal events from
$\nu_\mu \to \nu_\mu,\nu_e$
oscillations for \NBBn2, \NBBn3 and 
those from $\overline{\nu_\mu} \to \overline{\nu_\mu},\overline{\nu_e}$
oscillations for \NBBa2, with 1Mt$\cdot$year exposure.
The results are shown for the parameters of
\eqref{std_input}.
}}
\tblab{signals}
\end{table}

Typical numbers of expected CC signals 
are tabulated in \Tbref{signals} for
the parameter sets\footnote{{%
Recently KamLAND collaboration confirmed that only the LMA 
solution of the solar-neutrino deficit problem is consistent 
with the data\cite{KamLAND2002}.  The allowed region of 
$\msun{}$ is found to be either 
$(6-9)$ or $(13-19)\times 10^{-5}$eV$^2$, slightly below or 
above our input value.  The conclusions of this paper 
remain valid no matter which region its true value is.
}}:
\bseq
\begin{eqnarray}
\satms{}&=&0.5\,, ~~~~~~~~~~~~~~~
\delta m^2_{13}=
\matm{}=\numt{3.5}{-3}~{\rm eV}^2\,, 
\eqlab{std_input_atm}
\\
\ssun{}&=&0.8\,, ~~~~~~~~~~~~~~~
\delta m^2_{12}=
\msun{}=\numt{1.0}{-4}~{\rm eV}^2\,, 
\eqlab{std_input_sol}
\\
\schz{}&=&0.06,~0.01\,, 
~~~~~
\dmns{}=0^\circ\,,~90^\circ\,,~180^\circ\,,~270^\circ\,, 
\eqlab{std_input_delta}\\
\rho&=&3 ~{\rm g/cm^3}\,.
\eqlab{std_input_matter}
\end{eqnarray}
\eqlab{std_input}
\eseq
\noindent
The numbers in the \Tbref{signals} are
for 1 Mt$\cdot$year exposure 
with $10^{21}$ POT per year for 0.77 MW operation of
HIPA at $L=295$ km.
 From \Tbref{signals}, we learn that the transition events,
$N_e$ and $\overline{N^{}}_{\overline{e}}$,
are sufficiently large to have the potential of
distinguishing the CP conserved cases,
$\dmns{}=0^{\circ}$ and $180^{\circ}$,
from the CP violating cases of
$\dmns{}=90^{\circ}$ and $270^{\circ}$,
even if $\schz{}=0.01$.
 We also find that the survival events,
$N_\mu$ and $\overline{N^{}}_{\overline{\mu}}$,
barely depend on the CP phase.
 The ratio
$\overline{N^{}}_{\overline{\mu}}(2$GeV$)$/$N_\mu(2$GeV$)$ is
approximately
$\sigma_{\nu_{\mu}}^{\rm CC}/
\sigma_{\overline{\nu}_{\mu}}^{\rm CC}\simeq 2.9$,
because both the flux and the survival rates are
approximately the same for $\nu_\mu$ and $\overline{\nu}_\mu$
\cite{H2B}.
 From the comparison of $N_\ell(2$GeV$)$ and $N_\ell(3$GeV$)$,
we find that $N_\mu(3$GeV$)$/$N_\mu(2$GeV$)\sim 3$ because of
the rise in the cross section ($\sim 1.5$) and the increase
in the survival rate $(\sim 2)$. The $\nu_e$ appearance signal
$N_e$ increases only slightly at higher energies
because a slight decrease in the transition probability
cancels partially the effect of the rising cross section.
Most notably, we find that the difference between the predictions of
$\dmns{}=0^\circ$ and $180^\circ$ cases is significantly larger for
$N_e(\nu_\mu^{};\vev{p_\pi^{}}=3$GeV$)$ than that for 
$N_e(\nu_\mu^{};\vev{p_\pi^{}}=2$GeV$)$.

 The above results can be seen clearly in \Fgref{signals},
where we show the expected number of ${\overline{\nu}_e}$ CC events
$\ov N_{\ov e}$ for \NBBa{2} with 4Mton$\cdot$year plotted
against those of the $\nu_e^{}$ CC event
$N_e^{}$ for \NBBn2 (left) and for \NBBn3 (right),
both with 1Mton$\cdot$year.
The CP-phase dependence of the predictions are shown as closed circles
for the parameters of \eqref{std_input} at $\schz{}=0.06$, 0.04, 0.02,
and 0.01.
Comparable numbers of $\overline{\nu}_e$ CC events $(\overline{N}_{\ov e})$
and ${\nu}_e$ CC events $({N}_e)$ are expected by giving 4 times more
$\overline{\nu}_\mu$ than $\nu_\mu$ beams.
At each $\schz{}$ the $\nu_\mu^{} \to \nu_e^{}$ events are expected to be
smaller at $\dmns{}=90^\circ$ (solid-squares) than at 
$\dmns{}=270^\circ$ (open-squares).
The trend is opposite for the $\overline{\nu}_\mu \to \overline{\nu}_e$
events, and thus anti-correlation allows us to distinguish the two cases
clearly.
On the other hand, the expected number at 
$\dmns{}=0^\circ$ (solid-circles) and that
at $\dmns{}=180^\circ$ (open-circles) do not differ much for
\NBBn2 and \NBBa2.
 We find that \NBBn3 predicts significant differences between
the two CP-invariant cases without loosing event numbers.

\begin{figure}[tb]
\begin{center}
{\scalebox{0.35}{\includegraphics{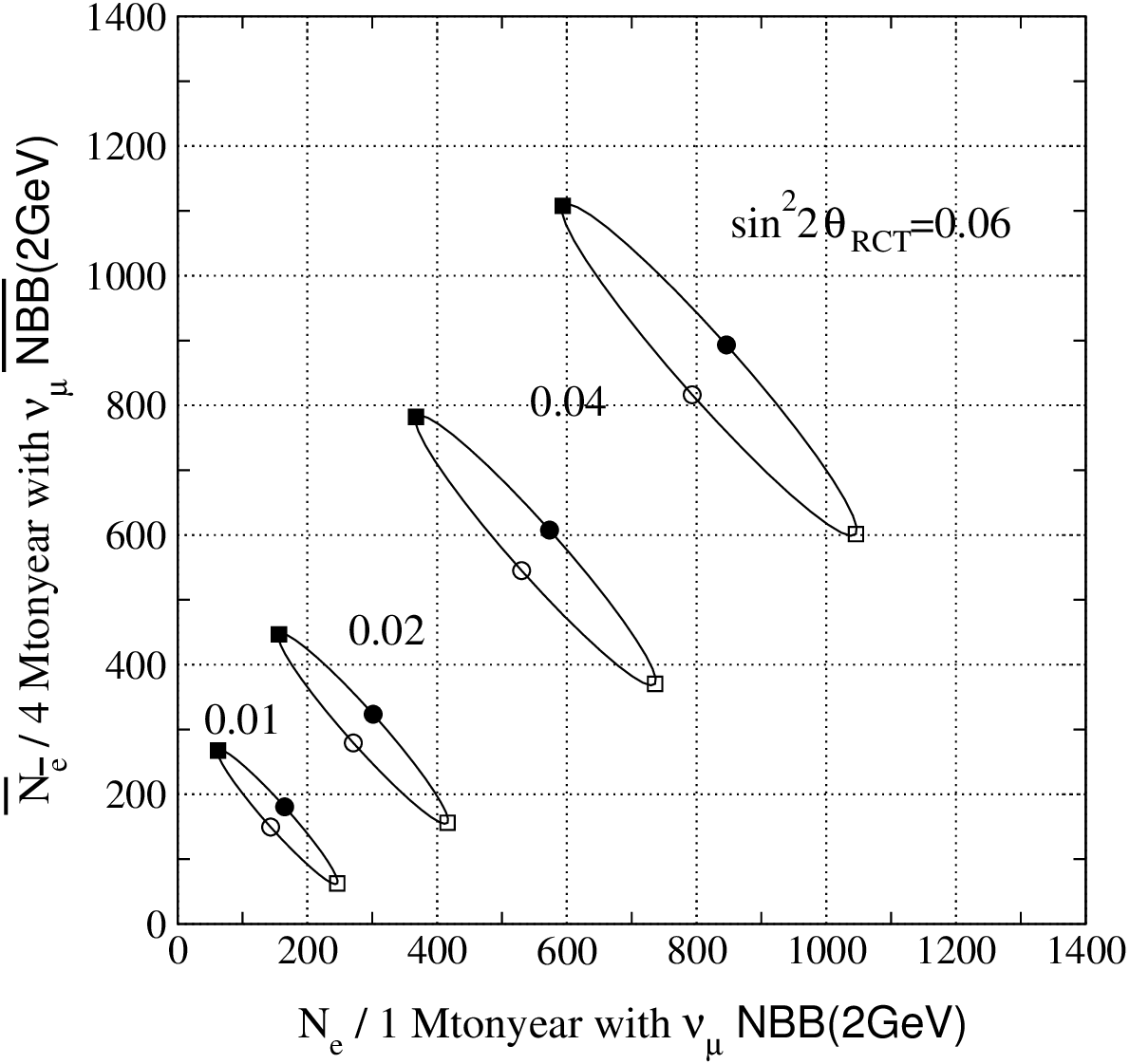}}} 
~~~
{\scalebox{0.35}{\includegraphics{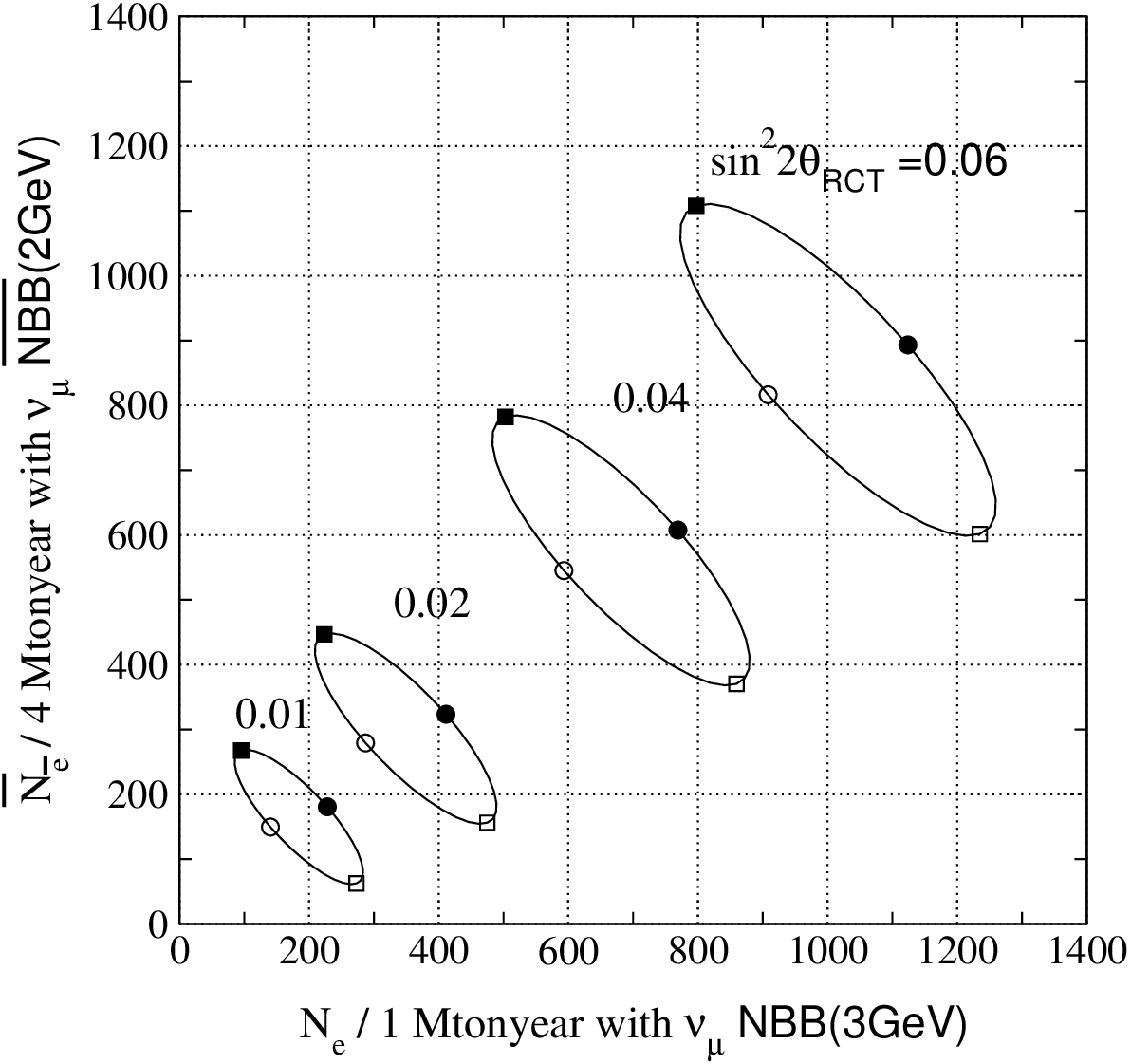}}} 
\end{center}	     
\caption{%
The CP phase dependence of
$N_{\overline{e}}(\overline \nu_\mu;2{\rm GeV})$ 
for 4 Mt$\cdot$year plotted against 
$\ov{N}_e(\nu_\mu;2{\rm GeV})$ 
for 1 Mt$\cdot$year in the left figure,
and against $N_{e}(\nu_\mu;3{\rm GeV})$ 
for 1 Mt$\cdot$year in the right figure.
$\dmns{}=0^\circ$ (solid-circle),
$\dmns{}=90^\circ$ (solid-square),
$\dmns{}=180^\circ$ (open-circle),
and
$\dmns{}=270^\circ$ (open-square).
The results are for
the parameters at \eqref{std_input}.
}
\Fglab{signals}
\end{figure}
In this report, we assume 1Mton$\cdot$year exposure each with
\NBBn2 and \NBBn3  and 4Mton$\cdot$year exposure of
\NBBa2, and examine the capability of HIPA-to-Hyper-Kamiokande
experiments to measure the CP phase, $\dmns{}$, under the
following simplified
treatments of the backgrounds and systematic errors.

For the $\nu_e^{}$ and $\nu_\mu^{}$ CC signal from
${\rm NBB}$($\nu_\mu^{};\vev{p_\pi^{}})$,
$N_e(\nu_\mu;\vev{p_\pi})$ and 
$N_\mu(\nu_\mu;\vev{p_\pi})$, respectively,
we consider the following backgrounds:
\bseq
\begin{eqnarray}
N_e(\vev{p_\pi^{}})_{\rm BG} &=&
 N_e(\nu_e^{};\vev{p_\pi^{}})
+N_{\overline{e}}(\overline\nu_\mu^{};\vev{p_\pi^{}})
+N_{\overline{e}}(\overline\nu_e^{}  ;\vev{p_\pi^{}})
+N_{e,\overline e}(NC;\vev{p_\pi^{}})\,,~~~~~~
\eqlab{BG_ne} \\
N_\mu(\vev{p_\pi^{}})_{\rm BG} &=&
 N_\mu(\nu_e^{};\vev{p_\pi^{}})
+N_{\overline{\mu}}(\overline\nu_\mu^{};\vev{p_\pi^{}})
+N_{\overline{\mu}}(\overline\nu_e^{}  ;\vev{p_\pi^{}})\,.
\eqlab{BG_nmu}
\end{eqnarray}
\eqlab{BG_n}
\eseq
\noindent
The first 3 terms in the r.h.s. are calculated as 
\begin{equation}
N_{\stackrel{(-)}{l}}(\stackrel{(-)}{\nu_\alpha^{}};\vev{p_\pi^{}})
=M N_A \int_{0}^{10~\mbox{\tiny GeV}}
dE_\nu
\Phi_{\stackrel{(-)}{\nu_\alpha^{}}}
(E_\nu;\vev{p_\pi^{}}) 
P_{\stackrel{(-)}{\nu_\alpha^{}} \to \stackrel{(-)}{\nu_l^{}}}(E_\nu)
\sigma^{\rm CC}_{\stackrel{(-)}{\nu_l^{}}}(E_\nu)\,,
\eqlab{def_ErrN}
\end{equation}
where $\Phi_{\nu_\alpha^{}}$ and $\Phi_{\overline \nu_\alpha^{}}$ stands,
respectively, for the secondary $\nu_\alpha^{}$ and $\overline\nu_\alpha^{}$
flux of the primarily $\nu_\mu^{}$ NBB.
The last term in \eqref{BG_ne} for the $e$-like events gives the
contribution of the NC
events where produced $\pi^0$'s mimic the electron
shower in the HK. By using the estimations from the
K2K experiments \cite{K2K}, we use
\begin{equation}
 N_{e,\overline e}(NC;\vev{p_\pi^{}}) =
P_{e/NC}~
\sum_{\nu_\alpha^{}=
\nu_e^{},\overline\nu_e^{},\nu_\mu^{},\overline\nu_\mu^{}}
N_{\nu_\alpha^{}}^{NC}(\vev{p_{\pi}^{}})\,,
\end{equation}
with
\begin{equation}
 P_{e/NC}^{} = 0.25\times(1 \pm 0.1)\%\,,
\eqlab{e_pi_err}
\end{equation}
where the NC event numbers are calculated as in 
\eqref{def_ErrN} by replacing $\sigma^{\rm CC}_{\nu_\ell}$
by $\sigma^{\rm NC}_{\nu_\ell}$.
The 10\% error in the misidentification probability of 0.25\% is
accounted for as a systematic error \cite{JHF2SK}.
The $\tau$-lepton contribution is found
to be negligibly small for the NBB's considered in this analysis.
The background for the $\overline{\nu}_\mu$ enriched beam \NBBa2
are evaluated in the same way.

Summing up, the event numbers for each energy neutrino and
anti-neutrino NBB's are calculated from the sum :
\bseq
\begin{eqnarray}
N_l^{}(\vev{p_\pi^{}}) &=&
 N_l^{}(\nu_\mu^{};\vev{p_\pi^{}})+
 N_l^{}(\vev{p_\pi^{}})_{\rm BG}\,, \\
\overline{N^{}}_{{l}^{}}(\vev{p_\pi^{}}) &=&
 \overline{N^{}}_{\ov{l}}(\overline{\nu_\mu^{}};\vev{p_\pi^{}})+
 \overline{N^{}}_{l}(\vev{p_\pi^{}})_{\rm BG}\,.
\end{eqnarray}
\eqlab{def_Nemu}
\eseq
\begin{table}[tb]
\begin{center}
\begin{small}
\begin{tabular}{|c||c|c|c|c|c|}
\hline
 NBB$(\vev{p_\pi})$ &
 & $\nu_{\mu}^{}$     &$\nu_{e}^{}$
 & $\overline \nu_{\mu}^{}$ &$\overline \nu_{e}^{}$  \\
\hline 
{${\rm NBB}$($2$GeV)}&
 CC &
 {\bf 2.8$\times$10}$^4$(1)&
 $\numt{2.2}{2}$(0.008)&
 $\numt{1.9}{2}$(0.007)&
 $\numt{1.3}{1}$(0.0005) \\
\cline{2-6}
 {1Mton$\cdot$year}&
 NC &
 {\bf 1.1$\times$10}$^4$(1)&
 $\numt{8.1}{1}$(0.007)&
 $\numt{8.1}{1}$(0.007)&
 $5.3$(0.0004) \\
\hline
{${\rm NBB}$($3$GeV)}&
 CC &
 {\bf 4.5$\times$10}$^4$(1)&
 $\numt{3.1}{2}$(0.007)&
 $\numt{2.0}{2}$(0.004)&
 $\numt{1.5}{1}$(0.0003) \\
\cline{2-6}
 {1Mton$\cdot$year}&
 NC&
 {\bf 1.6$\times$10}$^4$(1)&
 $\numt{1.1}{2}$(0.006)&
 $\numt{8.6}{1}$(0.005)&
 $6.3$(0.0004) \\
\hline
{$\overline{\rm NBB}$($2$GeV)}&
 CC &
 $\numt{3.0}{3}$(0.09)&
 $\numt{1.9}{2}$(0.005)&
 {\bf 3.5$\times$10}$^4$(1)&
 $\numt{2.5}{2}$(0.007) \\
\cline{2-6}
 {4Mton$\cdot$year}&
  NC &
 $\numt{1.2}{3}$(0.08)&
 $\numt{6.9}{1}$(0.005)&
 {\bf 1.5$\times$10}$^4$(1)&
 $\numt{1.0}{2}$(0.007) \\
\hline
\end{tabular}
\end{small}
\end{center}
\caption{%
Expected number of the CC and NC events at HK
in the absence of oscillations.
The results are for
1 Mton$\cdot$year
for the $\nu_\mu$ enriched
NBBs and 4Mton$\cdot$year for 
$\overline\nu_\mu$ enriched
NBB from HIPA.
The numbers in the parenthesis give
the fraction of each mode against the 
main mode whose numbers are shown by
bold letters.
}
\tblab{total_events}
\end{table}
\noindent
Most importantly,
we do not require the capability of the HK detector to distinguish
charges of electrons and muons.
In \Tbref{total_events}
the expected numbers of CC and NC events at HK
in the absence of oscillations are shown for
1Mton$\cdot$year each
for the $\nu_\mu$ enriched NBB's and
4Mton$\cdot$year for 
$\overline\nu_\mu$ enriched NBB.
The event numbers from the main (enriched) neutrinos
are shown by the bold letters.
The numbers in the parenthesis are the fractions as compared to the 
corresponding main mode.
From the comparison between \NBBn2 and \NBBa2,
we find that the fraction of the secondary-beam
contributions is much larger for the $\overline{\nu}_\mu$-beam
than that for the $\nu_\mu$-beam.
This is essentially because $\ov\nu_\ell^{}$
CC cross section is about a factor of three smaller than the
$\nu_\ell^{}$ CC cross section at
$E_\nu \sim 1$GeV.
\begin{figure}[tp]
\begin{center}
{\scalebox{0.25}{\includegraphics{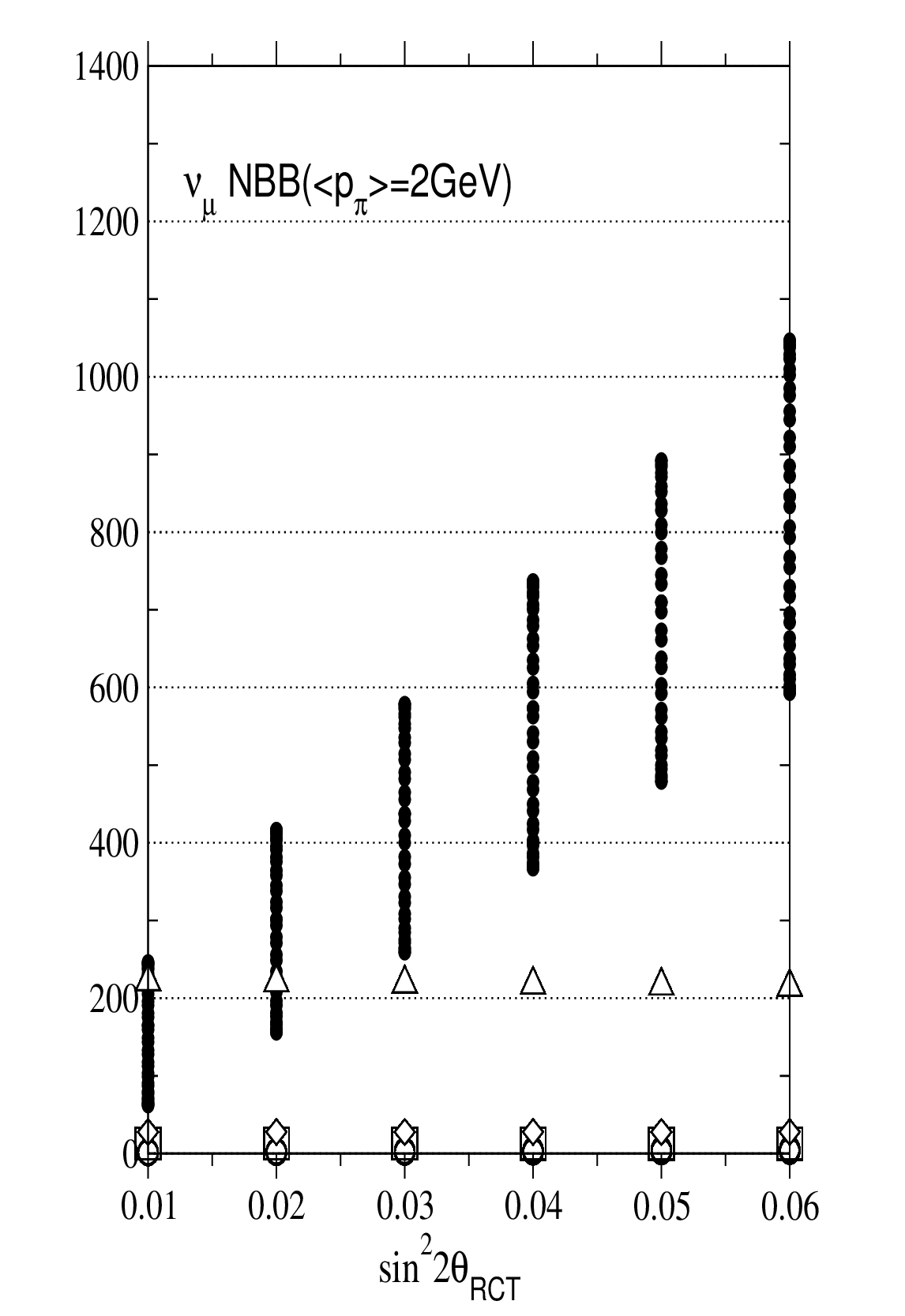}}} 
{\scalebox{0.25}{\includegraphics{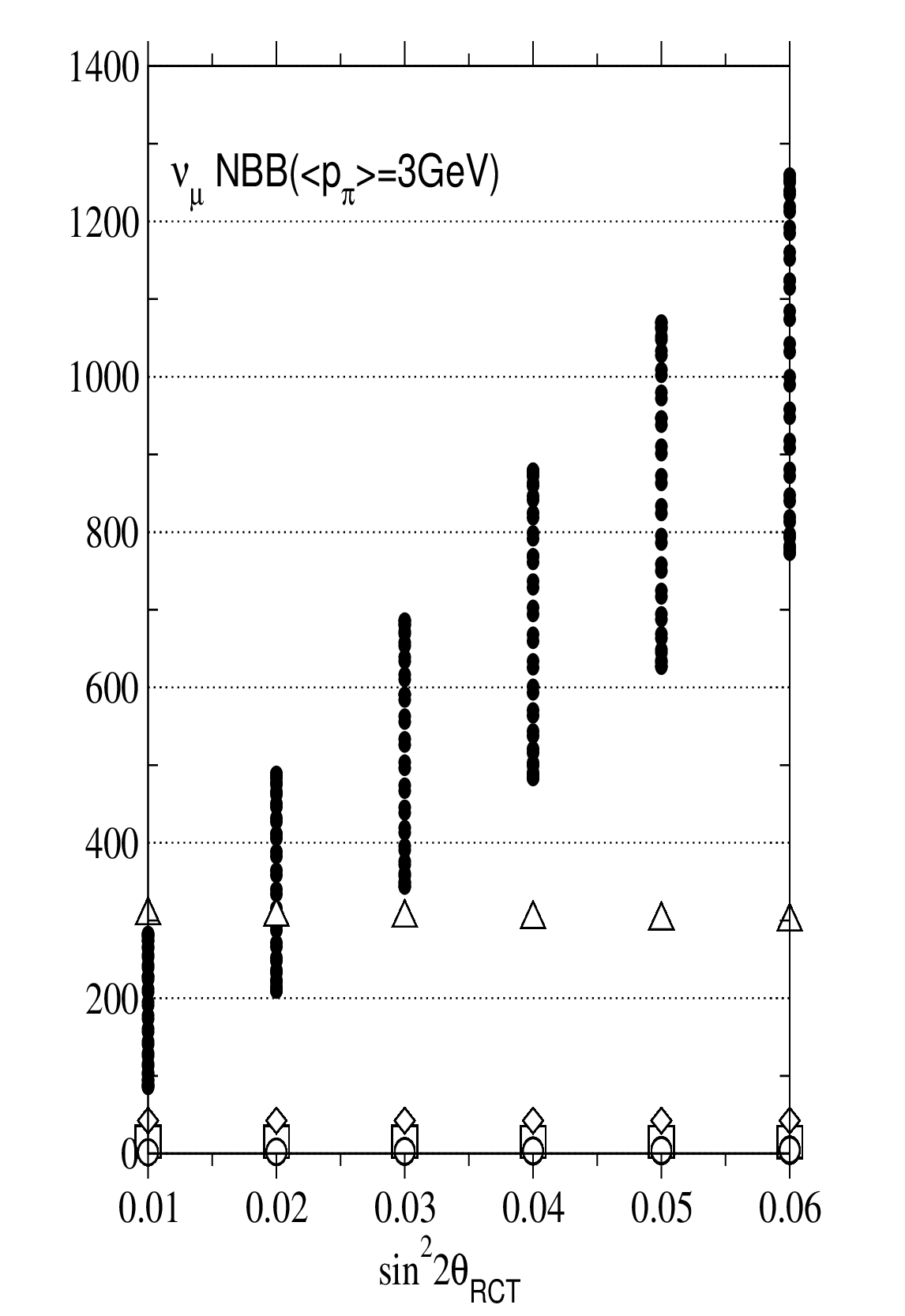}}} 
{\scalebox{0.25}{\includegraphics{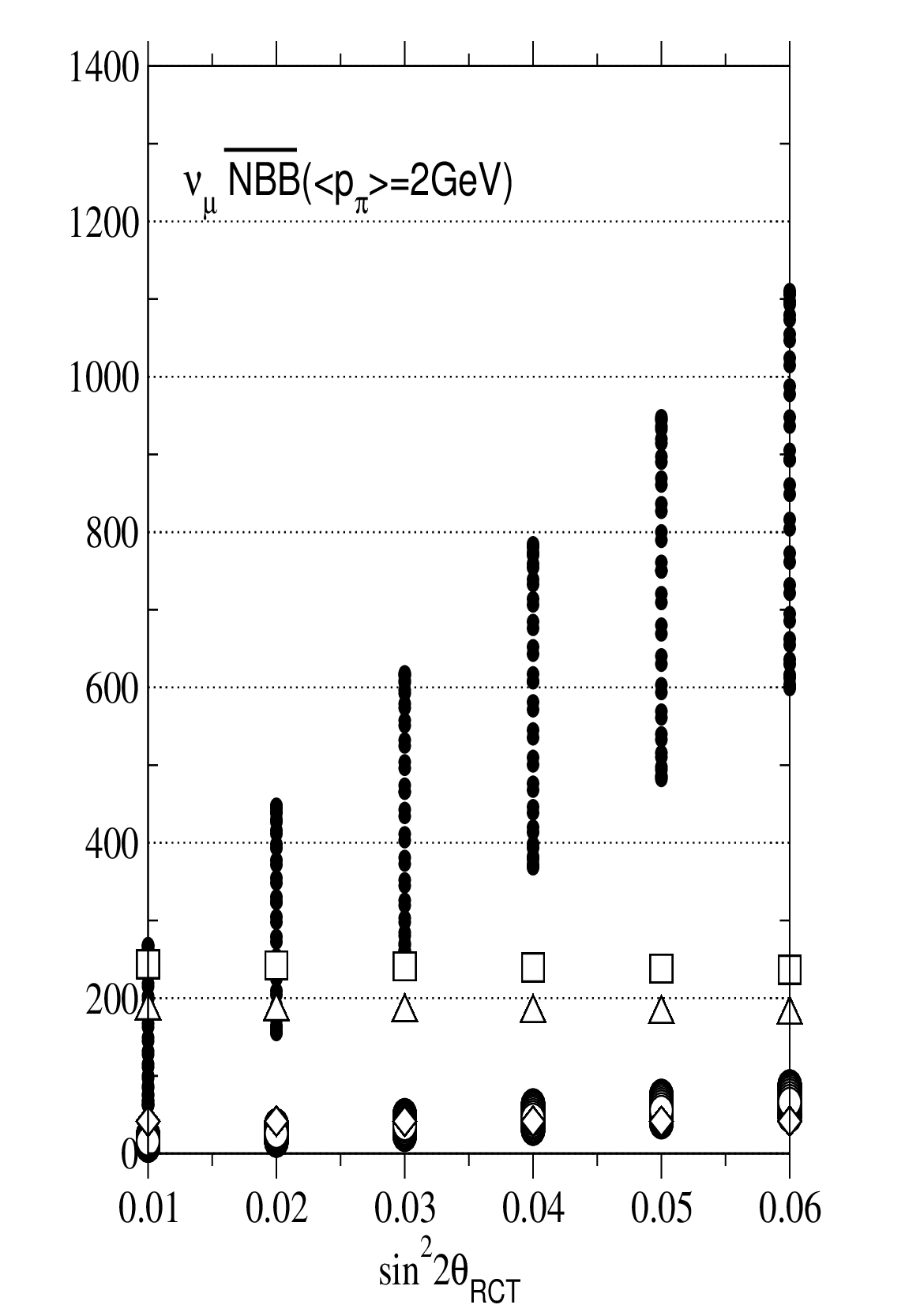}}} 
\caption{%
The $\schz{}$ dependence of the expected signal and background
event numbers for the parameters of
\eqref{std_input} for $\schz{}=0.01 \sim 0.06$.
Solid-circles stand for the number of expected signal events
for $\dmns{}=n\times 10^\circ$ $(n=1 \sim 36)$.
Open-diamonds denote the $\pi^0$ background from the NC events.
Open-triangles and open-square show $\nu_e$ and
$\overline \nu_e$ survival events.
Open-circles are
$\overline \nu_\mu \to \overline \nu_e$
transition events for \NBBn{2,3}
and
$\nu_\mu \to \nu_e$
transition events for \NBBa{2}.
}
\Fglab{bg}
\end{center}
\end{figure}

In \Fgref{bg}, we show the expected 
$\nu_\mu \to \nu_e$ ($\overline{\nu}_\mu \to \overline{\nu}_e$)
signal and background
event numbers for the parameters of \eqref{std_input}
for $\schz{}=0.01 \sim 0.06$.
 The solid-circles show the number of expected signal events
for $\dmns{}=n\times 10^\circ$ $(n=1 \sim 36)$.
 The numbers of signal events are largest at around $\dmns{}=270^\circ$
for \NBBn{2,3}, 
while those for \NBBa{2} are largest at around $90^\circ$,
as is expected from the CP phase dependence 
of $N_e$ and $\ov{N}_{\overline{e}}$ shown in \Fgref{signals}.
 The open-triangle denotes $\nu_e \to \nu_e$ CC events,
which give the largest background for the experiments with
\NBBn{2,3},
and the second largest background for \NBBa2.
 The open-square denotes $\ov\nu_e \to \ov\nu_e$ CC events
that gives the largest background for \NBBa2,
but is negligible for \NBBn{2,3}.
 The open-diamond denotes the background from the NC events,
where $\pi^0$'s are miss-identified as electrons.
 They give the second largest background for \NBBn{2,3}.
 Backgrounds from $\ov \nu_\mu \to \ov \nu_e$
transition events for \NBBn{2,3} and those from $\nu_\mu \to \nu_e$
transition events for \NBBa{2} are shown by open-circle.
 These transition backgrounds depend on the CP phase
and they tend to cancel the $\dmns{}$ dependence of the
signals,
but their magnitudes are small.
The background level starts dominating the signal
at $\schz{}\lsim 0.02$.

The background numbers for the $\mu$-like signals are found to be
negligibly small ($\sim 10^{-2}$) for \NBBn{2,3}.
Those for \NBBa{2} are found to be about 21\% of the signal 
almost independent of $\schz{}$.
In Both cases, the major background comes from the secondary 
$\ov\nu_\mu$ ($\nu_\mu$) survival events.

Our analysis proceeds as follows.
For a given set of the model parameters,
we calculate the expected numbers of all the signal
and background events for each NBB($\vev{p_\pi}$)
and $\ov{\rm NBB}(\vev{p_\pi})$,
by assuming 100\% detection efficiencies for simplicity.
The resulting numbers of $\mu$-like and $e$-like events
are then denoted by $N_\mu^{true}(\vev{p_\pi})$ and 
$N_e^{true}(\vev{p_\pi})$ for ${\rm NBB}(\vev{p_\pi})$,
and $\ov{N_\mu}^{true}(\vev{p_\pi})$ and
$\ov{N_e}^{true}(\vev{p_\pi})$ for $\ov{\rm NBB}(\vev{p_\pi})$.

We account for the following two effects
as major parts of the systematic uncertainty
in this analysis.
One is the uncertainty in the total flux of 
each neutrino beam,
for which we assign the uncertainty,
\begin{equation}
 \stackrel{(-)}f_{\nu_\alpha^{}}(\vev{p_\pi^{}})=1\pm0.03,
\eqlab{flux_un}
\end{equation}
independently for $\nu_\alpha^{}=\nu_e^{},\nu_\mu^{},\ov\nu_e^{},\ov\nu_\mu^{}$
and for \NBBn{2}, \NBBn{3}, and \NBBa{2}.
Although it is likely that correlation exists among the flux
uncertainties,
we ignore possible effects of correlations in this analysis.
By using the above flux factors,
theoretical predictions for the event numbers,
$N^{fit}_l(\vev{p_\pi^{}})$ and
$\overline{N}^{fit}_{l}(\vev{p_\pi^{}})$,
are calculated as
\bseq
\begin{eqnarray}
N^{fit}_\ell(\vev{p_\pi^{}})
&=&
f_{\nu_e^{}}(\vev{p_\pi^{}})
N_\ell(\nu_e^{},     \vev{p_\pi^{}})
+
f_{\nu_\mu^{}}(\vev{p_\pi^{}})
N_\ell(\nu_\mu^{},   \vev{p_\pi^{}})
\nn\\
&&+
f_{\overline\nu_e^{}}(\vev{p_\pi^{}})
N_{\overline{\ell}}(\overline\nu_e^{},  \vev{p_\pi^{}})
+
f_{\overline\nu_\mu^{}}(\vev{p_\pi^{}})
N_{\overline{\ell}}(\overline\nu_\mu^{},\vev{p_\pi^{}})\nn\\
&&
+\delta_{\ell,e}~P_{e/NC}~{}
\sum_{\nu_\alpha^{}}
f_{\nu_\alpha^{}}(\vev{p_\pi^{}})
N_{\nu_\alpha^{}}^{NC}(\vev{p_\pi^{}})\,,\\
\overline{N}^{fit}_\ell(\vev{p_\pi^{}})
&=&
\overline{f}_{\nu_e^{}}(\vev{p_\pi^{}})
\overline{N}_\ell(\nu_e^{},     \vev{p_\pi^{}})
+
\overline{f}_{\nu_\mu^{}}(\vev{p_\pi^{}})
\overline{N}_\ell(\nu_\mu^{},   \vev{p_\pi^{}})
\nn\\
&&+
\overline{f}_{\overline\nu_e^{}}(\vev{p_\pi^{}})
\overline{N}_{\overline{\ell}}(\overline\nu_e^{},  \vev{p_\pi^{}})
+
\overline{f}_{\overline\nu_\mu^{}}(\vev{p_\pi^{}})
\overline{N}_{\overline{\ell}}(\overline\nu_\mu^{},\vev{p_\pi^{}})\nn\\
&&+\delta_{\ell,e}~P_{e/NC}~{}
\sum_{\nu_\alpha^{}}
\overline{f}_{\nu_\alpha^{}}(\vev{p_\pi^{}})
\overline{N}_{\nu_\alpha^{}}^{NC}(\vev{p_\pi^{}})\,,
\end{eqnarray}
\eqlab{def_fit}
\eseq
\noindent
where the last terms proportional to $\delta_{\ell,e}$ are
counted only for $\ell=e$.
As the second major systematic error,
we allocate 3.3\% overall uncertainty in the matter density along
the base-line, \eqref{def_err_mat}. 
The fit functions are hence calculated for
an arbitrary set of the 6 model parameters, 
the 12 flux normalization factors,
and the matter density $\rho$.

The $\chi^2$ function of the fit in this analysis
can now be expressed as 
\begin{eqnarray}
\chi^2 &=& 
\sum_{NBB}
\left\{
\left(
\dfrac{{N^{fit}_{\mu}}(\vev{p_\pi}) -
       {N^{true}_{\mu}}(\vev{p_\pi})}
{{\sigma_\mu}(\vev{p_\pi})}
\right)^2
+
\left(
\dfrac{{N^{fit}_{e}}(\vev{p_\pi}) - 
       {N^{true}_{e}}(\vev{p_\pi})}
{\sigma_{e}(\vev{p_\pi})}
\right)^2
\right.\nn \\
&&
\hspace*{20ex}
+
\left.
\sum_{\nu_\alpha^{}}
\left(
\dfrac{f_{\nu_\alpha^{}}(\vev{p_\pi^{}})-1.0}{0.03}
\right)^2
\right\} \nn \\
&&+
\sum_{\ov{NBB}}
\left\{
\left(
\dfrac{{\overline{N}^{fit}_{\mu}}(\vev{p_\pi}) -
       {\overline{N}^{true}_{\mu}}(\vev{p_\pi})}
{{\sigma_{\mu}}(\vev{p_\pi})}
\right)^2
+
\left(
\dfrac{{\overline{N}^{fit}_{e}}(\vev{p_\pi}) - 
       {\overline{N}^{true}_{e}}(\vev{p_\pi})}
{\sigma_{e}(\vev{p_\pi})}
\right)^2
\right.\nn\\
&&
\hspace*{20ex}
+
\left.
\sum_{\nu_\alpha^{}}
\left(
\dfrac{\ov{f}_{\nu_\alpha^{}}(\vev{p_\pi^{}})-1.0}{0.03}
\right)^2
\right\} \nn \\
&&
+
\left(
{\dfrac{\rho-3.0^{^{}}}{0.1_{_{}}}}
\right)^2
+
\left(
{\dfrac{\msun{fit}-\msun{true}}{0.1\times\msun{true}}}
\right)^2
+
\left(
{\dfrac{\ssun{fit}-\ssun{true}}{0.06}}
\right)^2
\,,
\eqlab{def_chi}
\end{eqnarray}
where the summation is over \NBBn{2}, \NBBn{3} and \NBBa2. 
Even though we have only one $\ov{\rm NBB}$
in our analysis, we retain the summation symbol in
\eqref{def_chi} for the sake of clarity.
The last two terms are added because 
KamLAND experiment \cite{KAMLAND} will 
measure $\msun{}$ at 10\% level
and 
the solar neutrino experiments constrain $\ssun{}$ with 
the $1\sigma$ error of about 0.06
for the LMA parameters of \eqref{std_input}.
The individual error for each
$N_{\mu}(\vev{p_\pi^{}})$
($\overline{N}_{\mu}(\vev{p_\pi^{}})$)
is statistical only,
whereas the error for each
$N_e(\vev{p_\pi^{}})$
($\overline{N}_{e}(\vev{p_\pi^{}})$)
is a sum of the statistical errors
and the systematic error coming from the
10\% uncertainty in the $e/\pi^0$ misidentification
probability of \eqref{e_pi_err},
\bseq
\begin{eqnarray}
\sigma_\mu(\vev{p_\pi})&=&
\sqrt{N^{true}_\mu(\vev{p_\pi})}\,,\\
\sigma_e(\vev{p_\pi})&=&
\sqrt{
N^{true}_e(\vev{p_\pi})
+
\left(
0.1N^{true}_{e,\ov e}(NC;\vev{p_\pi})
\right)^2
}\,.
\end{eqnarray}
\eqlab{def_sigma}
\eseq
\noindent
The errors for the \NBBa2 case are calculated
similarly as above.

\begin{figure}[htbp]
\begin{center}
{\scalebox{0.80}{\includegraphics{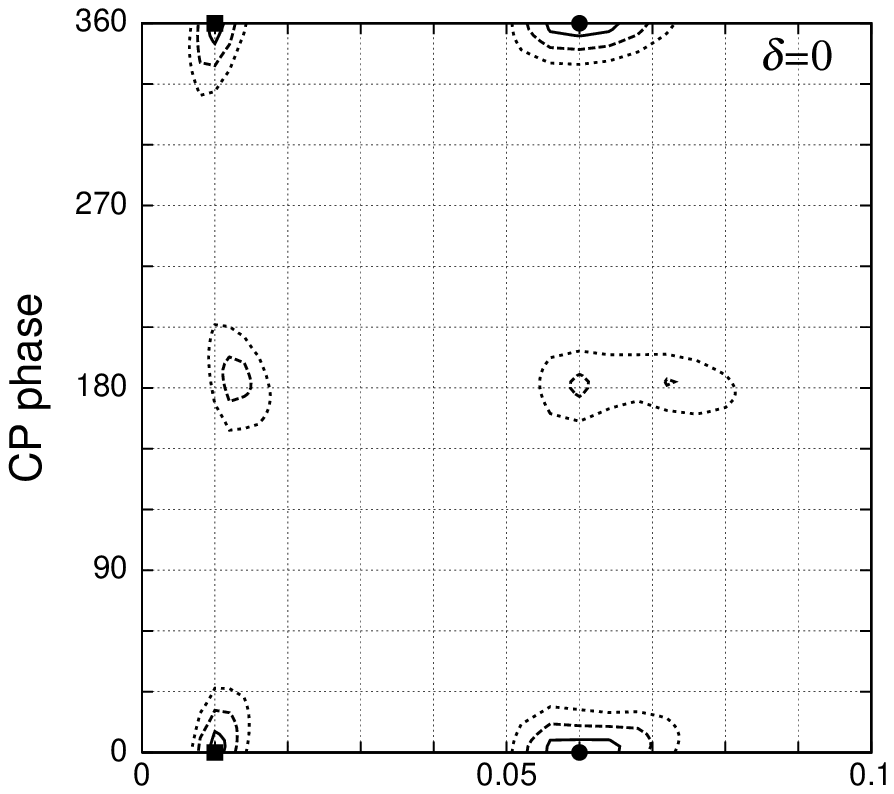}}} 
{\scalebox{0.80}{\includegraphics{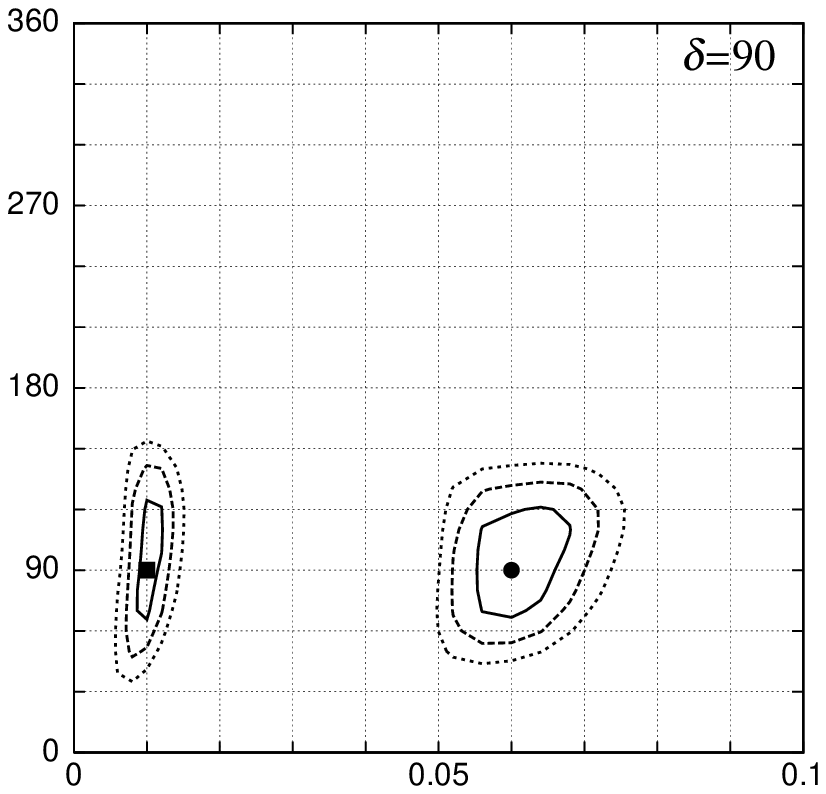}}} 
{\scalebox{0.80}{\includegraphics{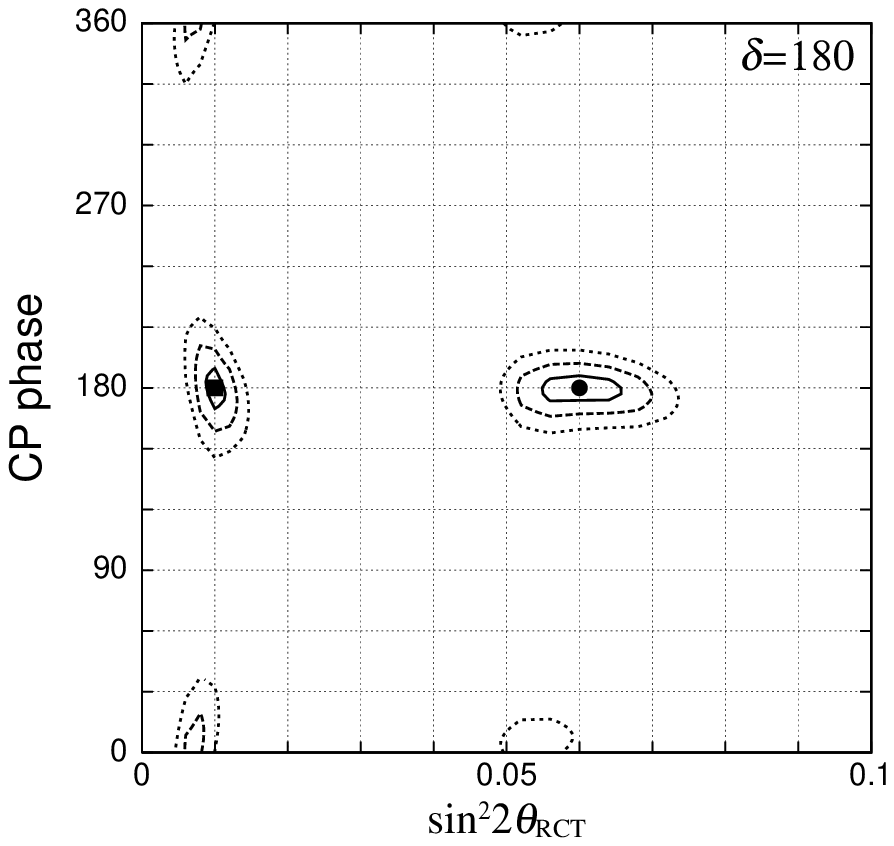}}} 
{\scalebox{0.80}{\includegraphics{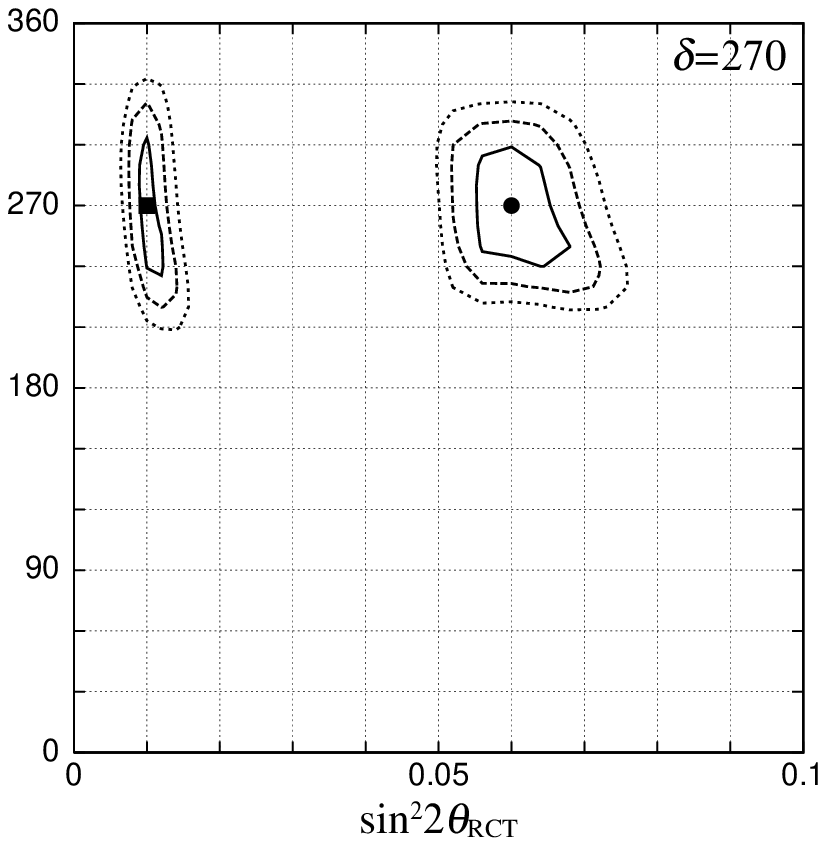}}} 
\end{center}	     
\caption{%
Regions allowed by the HIPA-to-HK experiment
are shown in the plain of $\schz{}$ and $\dmns{}$.
The assumed experimental conditions are 1 Mt$\cdot$year each
for \NBBn{2} and \NBBn{3}, and 4 Mt$\cdot$year for \NBBa2 with $10^{21}$
POT/year.
The input data are calculated for the LMA parameters of
\eqref{std_input}.
In each figure, the input parameter point
$(\schz{true}\,,\dmns{true})$ is shown by a solid-circle
for $\schz{true}=0.06$, and by a solid-square for 
$\schz{true}=0.01$. 
The regions where $\chi^2_{min}<$1, 4, and 9 are
depicted by solid, dashed, and dotted boundaries, respectively.
}
\Fglab{chi}
\end{figure}

\begin{figure}[htbp]
\begin{center}
{\scalebox{0.80}{\includegraphics{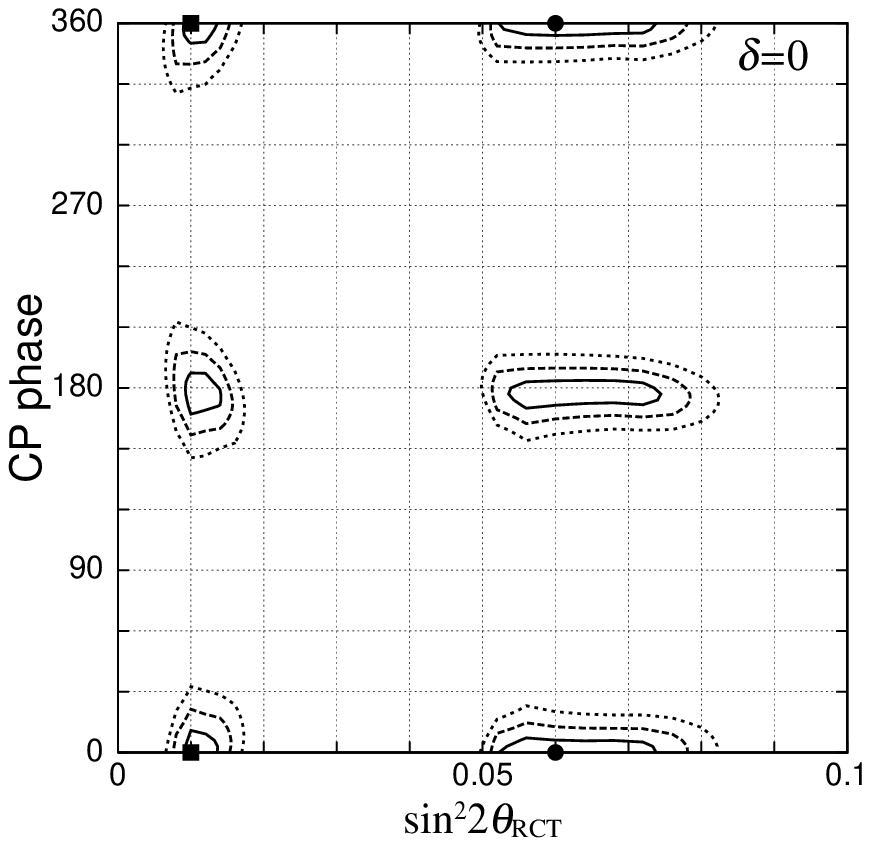}}} 
{\scalebox{0.80}{\includegraphics{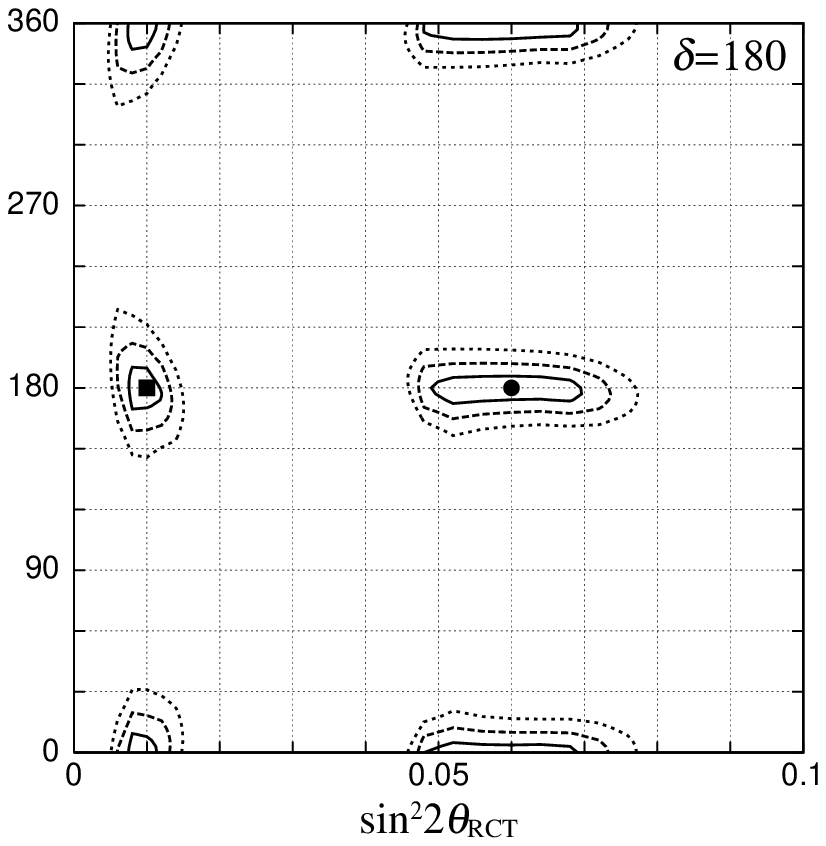}}} 
\end{center}	     
\caption{%
The same as \Fgref{chi} but with 2 Mt$\cdot$year for
\NBBn{2} and 4 Mt$\cdot$year for \NBBa{3} only.
The two-fold ambiguity in the fit is clearly seen.
}
\Fglab{chi2}
\end{figure}

We show in \Fgref{chi} 
regions allowed by the HIPA-to-HK experiment
in the plain of $\schz{}$ and $\dmns{}$.
The mean values of the input data are calculated for
the LMA parameters of \eqref{std_input}.
In each figure, the input parameter point
$(\schz{true}\,,\dmns{true})$ is shown by a solid-circle
for $\schz{true}=0.06$, and by a solid-square for 
$\schz{true}=0.01$. 
The regions where $\chi^2_{min}<$1, 4, and 9 are
depicted by solid, dashed, and dotted boundaries, respectively.
%The $\chi^2$ fit has been performed by restricting
%the solar-neutrino oscillation amplitude in the range
%\begin{eqnarray}
%0.7 \leq 
%\ssun{fit}
% \leq 0.9 \,,
%\eqlab{fit_sol}
%\end{eqnarray}
%which represents the allowed range of the LMA solution at
%present \cite{solar}.
All the 6 parameters,
$\matm{fit}$, 
$\satms{fit}$,
$\msun{fit}$,
$\ssun{fit}$
$\schz{fit}$,
$\dmns{fit}$,
the matter density
$\rho^{fit}$,
and the 12 flux normalization factors are allowed
to vary freely in the fit.

From the top-right and bottom-right figures
for $\dmns{true}=90^\circ$ and $270^\circ$ respectively,
we learn that $\dmns{}$ can be constrained to $\pm 30^\circ (\pm 60^\circ)$
at the 1$\sigma$ (3$\sigma$) level,
even if $\schz{true}=0.01$.
This is because $N_e + \ov N_e$ constrain $\schz{}$ and
$N_e/\ov N_e$ distinguishes between $\dmns{}=90^\circ$
and $270^\circ$ in \Fgref{chi}, whereas the remaining parameters
($\matm{}$ and $\satms{}$) are constrained by the $\nu_\mu$ and
$\ov\nu_\mu$ survival data, $N_\mu$ and $\ov N_\mu$.
The accuracy of the $\dmns{}$ measurement does not decrease
significantly for $\schz{true}=0.01$ despite the large background 
level,
because the $\dmns{}$-dependence of the signal
exceeds significantly the 3\% uncertainty
of the background level from the flux normalization factors
in \eqref{flux_un}.
We find that the CP violation signal
can be distinguished from the CP-conserving
cases ($\dmns{}=0^\circ$ or $180^\circ$) at 4$\sigma$ (3$\sigma$)
level for all $\dmns{}$ values in the region $|\dmns{}|$,
$|\dmns{}-180^\circ|>30^\circ$ if $\schz{true}\gsim 0.03$ (0.01),
for the LMA parameters of \eqref{std_input} and
for the systematic errors assumed in this analysis.

The situation is quite different for the CP-conserving cases of
$\dmns{true}=0^\circ$ or $180^\circ$ shown in the left-hand side
of \Fgref{chi}.
$\dmns{}$ can be constrained 
to better than $\pm 7^\circ$ ($11^\circ$) accuracy
at 1$\sigma$ level for $\schz{}\gsim0.06$ (0.01),
but the two cases cannot be distinguished at 2$\sigma$ level.
This is mainly because of the similarity of $N_e/\ov N_e$ 
between $\dmns{}=0^\circ$ and $180^\circ$ in \Fgref{signals}.
The difference between the two cases is larger for \NBBn3.
If we remove the \NBBn3 data from the fit, we find
that the two cases cannot be distinguished even at 1$\sigma$ level.
This two-fold ambiguity between $\dmns{}$ and $180^\circ-\dmns{}$
is found in general for all $\dmns{}$,
because the difference in the predictions can
be adjusted by a shift in the fitted $\schz{}$ value;
see \Fgref{signals}.

As a demonstration of the effect of using two NBB's, \NBBn{2} 
and \NBBn{3}, in the analysis, we show in \Fgref{chi2} the fit results 
when the data are generated by using \NBBn{2} and \NBBa{2} 
only, each at 2 Mton$\cdot$year and 4 Mton$\cdot$year, respectively.
It is clearly seen from the figures that the `mirror' solution at 
$\dmns{}=180^\circ$ $(0^\circ)$ can fit the data as well as the 
`true' solution at $\dmns{}=0^\circ$ $(180^\circ)$.  Essentially the 
same results are obtained when we replace \NBBn{2} by \NBBn{3} 
in the above analysis.  It is only by combining the two NBB's 
that we can distinguish the two solutions as shown in \Fgref{chi}.  
We find less significant difference from the results of \Fgref{chi} 
when the input $\dmns{}$ value is $90^\circ$ or $270^\circ$.

It is remarkable that the $1\sigma$ error of $\dmns{}$ is as large as
$30^{\circ}$ for $\dmns{true}=90^\circ$ and $270^\circ$
while it is less than $10^\circ$ for $\dmns{true}=0^\circ$ and
$180^\circ$.
This is simply because the $\dmns{}$ dependence of the 
$\nu_\mu$-to-$\nu_e$ (and also $\ov{\nu}_\mu$-to-$\ov{\nu}_e$) 
oscillation probability is roughly proportional to $\sin\dmns{}$, 
in the vicinity of the first dip of the $\nu_\mu$-to-$\nu_\mu$ 
survival probability.  
%The large $1\sigma$ error of the CP-violating cases reflects
%the uncertainty of $\ssun{}$,
%where we give the equal probability for any values of $\ssun{fit}$
%in the allowed range of \eqref{fit_sol}.
%The $\dmns{}$ dependence of the CP-violating cases come from 
%the Jarlskog parameter \cite{jar},
%which depend on the product of $\ssun{}$ and $\sin \dmns{}$.
%The 10\% uncertainty in $\ssun{}$ gives rise to about $20^\circ$
%uncertainty in $\dmns{}$.
%This leads to the large $1\sigma$ error of $\dmns{}$ around
%$\dmns{true}=90^\circ$ and $270^\circ$.
%When $\dmns{true}=0^\circ$ or $180^\circ$, the Jarlskog parameter
%vanishes and the uncertainty in $\ssun{}$ does not affect the error
%of $\dmns{}$.

We close this article by pointing out that the low-energy LBL experiment
like HIPA-to-HK cannot distinguish between the neutrino-mass hierarchy
cases (between I and III) because of the small matter effect at low energies.
If we repeat the analysis by using the same input data but 
assuming the hierarchy III in the analysis,
we obtain another excellent fit to all the data
where the fitted model parameters are slightly shifted from their
true (input) values.
VLBL experiments at higher energies at $L>1000$km \cite{H2B}
are needed to determine the mass hierarchy.

\noindent
{\it Acknowledgments}\\
The authors wish to thank stimulating discussions with
T.~Kajita, T.~Kobayashi and J.~Sato.
The work of MA is supported in part by
the Grant-in-Aid for Scientific Research from MEXT, Japan.
KH would like to thank the core-university
program of JSPS for support.
The work of NO is supported in part by a grand from 
the US Department of Energy, DE-FG05-92ER40709.

\end{document}